\documentstyle[12pt,epsfig]{article}
\def\beq{\begin{equation}}
\def\eeq{\end{equation}}
\def\beqa{\begin{eqnarray}}
\def\eeqa{\end{eqnarray}}

\def\pr{{\it Phys. Rev.}\ }
\def\prl{{\it Phys. Rev. Lett.}\ }
\def\pl{{\it Phys. Lett.}\ }

\def\mpl{{\it Mod. Phys. Lett.}\ }
\def\ijmp{{\it Int. Journ. Mod. Phys.}\ }

\def\cqg{{\it Class. Quantum Grav.}\ }
\def\aph{{\it Ann. Phys.}\ }

\def\grg{{\it Gen. Relativ. Grav.}\ }

\def\apj{{\it Ap. J.}\ }

\begin{document}
\def\bib#1{[{\ref{#1}}]}
\begin{titlepage}
\title{Curvature quintessence matched with observational data}

\author{{S. Capozziello\thanks{capozziello@sa.infn.it}, V. F. Cardone\thanks{winny@na.infn.it}, S. Carloni\thanks{carloni@sa.infn.it},
A. Troisi\thanks{antro@sa.infn.it}}
\\ {\it Dipartimento di Fisica``E. R. Caianiello'', } \\
 {\it Universit\`{a} di Salerno, I-84081 Baronissi, Salerno,} \\
 {\it Istituto Nazionale  di Fisica Nucleare, sez. di Napoli},\\{\it Gruppo
Collegato di Salerno,}\\{ Via S. Allende-84081 Baronissi (SA),
Italy}}

\date{\today}

\maketitle

\begin{abstract}
Quintessence issues can be achieved by taking into account higher
order curvature invariants into the effective action of
gravitational field. Such an approach is naturally related to
fundamental theories of quantum gravity which predict higher order
terms in loop expansion of quantum fields in curved space-times.
In this framework, we obtain a class of cosmological solutions
which are fitted against cosmological data. We reproduce
encouraging results able to fit high redshift supernovae and WMAP
observations. The age of the universe and other cosmological
parameters are discussed in this context.

\end{abstract}

\thispagestyle{empty} \vspace{20.mm}
 PACS number(s): 98.80.Cq, 98.80. Hw, 04.20.Jb, 04.50 \\

\vspace{5.mm}

\vfill

\end{titlepage}

\section{Introduction}

The existence of a dark energy term into cosmological dynamics has
become a paradigm in the last five years since several
observational campaigns give reliable indications on the apparent
acceleration of the universe.
\\
High redshift supernovae surveys \cite{perlmutter,riess}, CMBR
data \cite{boomerang,maxima,cobe}, Sunyaev-Zeldovich / X-ray
methods \cite{sunyaev} and other approaches provide a new picture
of the universe. It can be represented as an isotropic,
homogeneous, spatially flat 4-dim manifold filled with about 30\%
of baryonic and non-baryonic matter and about 70\% of dark energy,
simply referred as ``cosmological component".
\\
Also the very recent WMAP data \cite{wmap} confirm such a picture
with extremely low uncertainties in the estimate of cosmological
parameters.
\\
It is evident that the cosmological component should be the
ingredient capable of generating the accelerated expansion, but,
till now, its real nature is a puzzle which seems far to be solved.\\
Many approaches has been developed. Cosmological constant is the
most straightforward candidate, however it is ruled out since its
observational value (constrained by observations) differs of 120
order of magnitude from the theoretical prediction of QCD
\cite{carroll,starobinsky,straumann}. This argument does not allow
to interpret cosmological constant as the vacuum energy of
gravitational field unless an evolutionary mechanism is invoked in
order to explain dynamics starting from the huge early values of
energy ({\it cosmological constant problem}). Besides, the
observed comparable amounts (in order of magnitude) of matter and
dark energy sets a strong fine-tuning problem which cannot be
overcome considering wide ranges of initial
data ({\it coincidence problem}) \cite{steinhardt}.\\
A second approach is to consider the cosmological component as a
dynamical term. This scheme, usually called {\it quintessence},
can be achieved adding scalar fields into Einstein gravity. Such
scalar fields are assumed rolling their interaction potential
\cite{steinhardt,rubano}, as in the case of inflation but in a
very different energetic regime (today instead of early universe).
However major shortcomings come out due, essentially, to the {\it
ad hoc} forms of self-interaction potential. Several forms of
potential (inverse power law, exponential, etc,
\cite{starobinsky,steinhardt}) achieve quintessence prescriptions
(e.g. accelerated behaviour, $\Omega_{\Lambda}\simeq{0.7}$,
coincidence problem, etc.) but none of them seems to be directly
related to some fundamental quantum field theory. The situation
is, somehow, similar to that of inflationary cosmology: inflation
is a good paradigm but no single
model matches all requests.\\
On the other hand, alternative approaches can be pursued starting
from some fundamental theory \cite{sahni,cardassian,chaplygin}.
These schemes aim to improve the quintessence approach overcoming
the problem of scalar field potential, generating a dynamical
source for dark energy as an intrinsic feature. The goal would be
to obtain a comprehensive model capable of linking the picture of
early universe to the today observed one; that is, a model derived
from some effective theory of quantum gravity which, through an
inflationary period results in the today accelerated Friedmann
expansion driven by some $\Omega_{\Lambda}$-term.\\
From this point of view, theories including torsion
\cite{stornaiolo,capoz-torsione,quint-torsione} or higher order
curvature invariants \cite{capoz-quint curv} naturally come into
the game. In fact, the former ones allow to include spin matter
fields at a fundamental level in General Relativity, while the
latter ones come out in every quantization scheme of matter fields
in curved space-time \cite{starobinsky,birrell,buchbinder}.
Specifically higher-order terms in curvature invariants as $R^{2},
R_{\mu\nu}R^{\mu\nu},R \Box{R}$ are inescapable if we want to
obtain an effective action of gravity closed to the Planck epoch.
This scheme naturally give rise to inflationary behaviours
\cite{starobinsky80,kerner1,kerner2}.\\
In \cite{capoz-quint curv}, it is investigated the possibility
that such terms could act also today as a non-clustered form of
dark-energy, providing accelerated behaviours. Besides, other
authors have recently discussed curvature self-interactions of
cosmic fluid to get cosmological constant
\cite{pavon}.\\
In this paper, we want better to detail such an approach. We
provide exact solutions and, following the scheme in
\cite{quint-torsione}, we try to obtain observational constraints
on the so called {\it Curvature Quintessence} in order to see if
curvature contributions can actually match the data of recent
surveys.\\
The paper is organized as follows. In Sec.2, we summarize the
curvature quintessence approach and derive a class of exact
cosmological solutions. Sect.3 is devoted to the matching with
observational data. In particular, we fit our solutions against
the SNIa data and derive the age of the universe. A further
discussion is carried out in Sec.4 considering the very recent
WMAP data, which seem to better constrain the parameters of
cosmological solutions. Sec.5 is devoted to the conclusions.\\

\section{Curvature Quintessence}

 A generic fourth--order theory of gravity, in
four dimensions, is given by the action \cite{capoz-quint curv},
 \begin{equation}\label{3}
 {\cal A}=\int d^4x \sqrt{-g} \left[f(R)+{\cal L}_{(matter)} \right]\,{,}
 \end{equation}
 where $f(R)$ is a function of Ricci scalar $R$ and ${\cal L}_{(matter)}$
 is the standard matter Lagrangian density.
We are using physical units $8\pi G_N=c=\hbar=1$. The field
equations are

 \begin{equation}\label{4}
 f'(R)R_{\alpha\beta}-\frac{1}{2}f(R)g_{\alpha\beta}=
 f'(R)^{;\mu\nu}(g_{\alpha\mu}g_{\beta\nu}-g_{\alpha\beta}g_{\mu\nu})+ \tilde{T}^{(matter)}_{\alpha\beta}\,,
 \end{equation}
 which can be recast in the more expressive form
 \begin{equation}\label{5}
 G_{\alpha\beta}=R_{\alpha\beta}-\frac{1}{2}g_{\alpha\beta}R=T^{(curv)}_{\alpha\beta}+T^{(matter)}_{\alpha\beta}\,,
 \end{equation}
 where an stress-energy tensor has been defined for the
 curvature contributes
 \begin{equation}
 \label{6}
T^{(curv)}_{\alpha\beta}=\frac{1}{f'(R)}\left\{\frac{1}{2}g_{\alpha\beta}\left[f(R)-Rf'(R)\right]+
f'(R)^{;\mu\nu}(g_{\alpha\mu}g_{\beta\nu}-g_{\alpha\beta}g_{\mu\nu})
\right\}
 \end{equation}
 and
 \begin{equation}
 \label{7}
 T^{(matter)}_{\alpha\beta}=\frac{1}{f'(R)}\tilde{T}^{(matter)}_{\alpha\beta}\,,
 \end{equation}
 is the stress-energy tensor of matter. We have taken into account
 the nontrivial coupling to geometry; prime means the derivative with respect to $R$.
If $f(R)=R+2\Lambda$, we recover the standard second--order gravity.\\
In a Friedmann-Robertson-Walker (FRW) metric, the action (\ref{3})
reduces to the point-like one:
 \begin{equation}\label{8}
 {\cal A}_{(curv)}=\int dt \left[{\cal L}(a, \dot{a}; R, \dot{R})\,+{\cal
 L}_{(matter)}\right]
 \end{equation}
where the dot means the derivative with respect to the cosmic
time. In this case the scale factor $a$ and the Ricci scalar $R$
are the
canonical variables \cite{lambda,capoz-lambiase}. \\
It has to be stressed that the definition of $R$ in terms of $a,
\dot{a}, \ddot{a}$ introduces a constraint in the action (\ref{8})
\cite{capoz-quint curv}, by which we obtain by the Lagrange
multiplier technique the lagrangian
$$
 {\cal L}={\cal L}_{(curv)}+{\cal L}_{(matter)}=a^3\left[f(R)-R
 f'(R)\right]+6a\dot{a}^2f'(R)+
$$
\begin{equation}
\label{10} \qquad +6a^2\dot{a}\dot{R}f''(R)-6ka
 f'(R)+a^3p_{(matter)}\,,
 \end{equation}
 (the standard fluid matter
 contribution acts essentially as a pressure term
 \cite{quartic}).
 The Euler-Lagrange equations coming from (\ref{10}) give
 the second order system:
\begin{equation}
\label{11}
2\left(\frac{\ddot{a}}{a}\right)+\left(\frac{\dot{a}}{a}\right)^2+
\frac{k}{a^2}=-p_{(tot)}\,,
 \end{equation}
and
\begin{equation}
\label{12}
f''(R)\left\{R+6\left[\frac{\ddot{a}}{a}+{\left(\frac{\dot{a}}{a}\right)}^2+\frac{k}{a^2}\right]\right\}=0\,,
\end{equation}
constrained by the energy condition
\begin{equation}
\label{13}
 \left(\frac{\dot{a}}{a}\right)^2+\frac{k}{a^2}=\frac{1}{3}\rho_{(tot)}\,.
\end{equation}

\noindent Using Eq.(\ref{13}), it is possible to write down
Eq.(\ref{11}) as
 \begin{equation}
 \label{14}
 \left(\frac{\ddot{a}}{a}\right)=-\frac{1}{6}\left[\rho_{(tot)}+3p_{(tot)}
 \right]\,.
 \end{equation}
 The accelerated or decelerated behaviour of the scale factor
 depends on the r.h.s. of (\ref{14}). The accelerated behaviour is
 achieved if

\begin{equation}
\label{15} \rho_{(tot)}+ 3p_{(tot)}< 0\,.
\end{equation}

  \noindent To understand
 the actual effect of these terms, we can distinguish between the matter and the geometrical
 contributions
 \begin{equation}
 \label{16}
 p_{(tot)}=p_{(curv)}+p_{(matter)}\;\;\;\;\;\rho_{(tot)}=\rho_{(curv)}+\rho_{(matter)}\, .
 \end{equation}
 Assuming that all matter components have non-negative pressure, Eq.(\ref{15})
becomes:
\begin{equation}
\label{17} \rho_{(curv)}> \frac{1}{3}\rho_{(tot)}\,.
\end{equation}
The curvature contributions can be specified by considering the
stress-energy tensor (\ref{6}); we obtain a curvature pressure
\begin{equation}
\label{18}
p_{(curv)}=\frac{1}{f'(R)}\left\{2\left(\frac{\dot{a}}{a}\right)\dot{R}f''(R)+\ddot{R}f''(R)+\dot{R}^2f'''(R)
-\frac{1}{2}\left[f(R)-Rf'(R)\right] \right\}\,,
 \end{equation}
and a curvature energy-density:
\begin{equation}
\label{19}
\rho_{(curv)}=\frac{1}{f'(R)}\left\{\frac{1}{2}\left[f(R)-Rf'(R)\right]
-3\left(\frac{\dot{a}}{a}\right)\dot{R}f''(R) \right\}\, ,
 \end{equation}
 which account for the geometrical contributions into the thermodynamical variables.
 \\
 It is clear that the form of $f(R)$ plays an essential role for this model. We choose the $f(R)$ function as a generic power law of the scalar
curvature and we ask for power law solutions of the scale
factor. \\
Summarizing, we consider:
 \begin{equation}
 \label{22}
 f(R)=f_0 R^n\,,\qquad
 a(t)=a_0\left(\frac{t}{t_0}\right)^{\alpha}\,.
 \end{equation}
 The interesting cases are for $\alpha\geq 1$
 which give rise to accelerated expansion.\\
 Let us now concentrate now concentrate on the case with
 $\rho_{(matter)}=0$.
Inserting Eqs.(\ref{22}) into the above dynamical system, for a
spatially flat space-time ($k=0$), we obtain an algebraic system
for the parameters $n$ and $\alpha$

 \begin{equation} \label{27}
 \left\{ \begin{array}{ll} \alpha [\alpha(n-2)+2n^{2}-3n+1]=0 \\
 \\
\alpha[n^{2}+\alpha(n-2-n-1)]=n(n-1)(2n-1)\\
\end{array}
\right.
\end{equation}

\vspace{7mm}

\noindent from which the allowed solutions:
\begin{equation}
\label{28}
\begin{array}{cc} \alpha=0\,\, \rightarrow\,\, n=0,\,\,1/2,\,\,1\\ \\
\alpha=\displaystyle\frac{2n^2-3n+1}{2-n}\,,\,\, \forall{n}\,\
\,{\rm but}\,\ \, n\neq{2}\,.
\end{array}
\end{equation}
\vspace{7mm}

 The solutions with $\alpha=0$ are not interesting since
they provide static cosmologies with a non evolving scale
factor\footnote{This result match with the standard General
Relativity case (n=1) in absence of matter.}. On the other hand,
the cases with generic $\alpha$ and $n$ furnish an entire family
of significative cosmological models. By the plot in Fig.1 we see
that such a family of solutions admit negative and positive values
of $\alpha$ which give rise to accelerated behaviours.
\\
\begin{figure*}
\begin{center}
  \includegraphics[width=10cm]{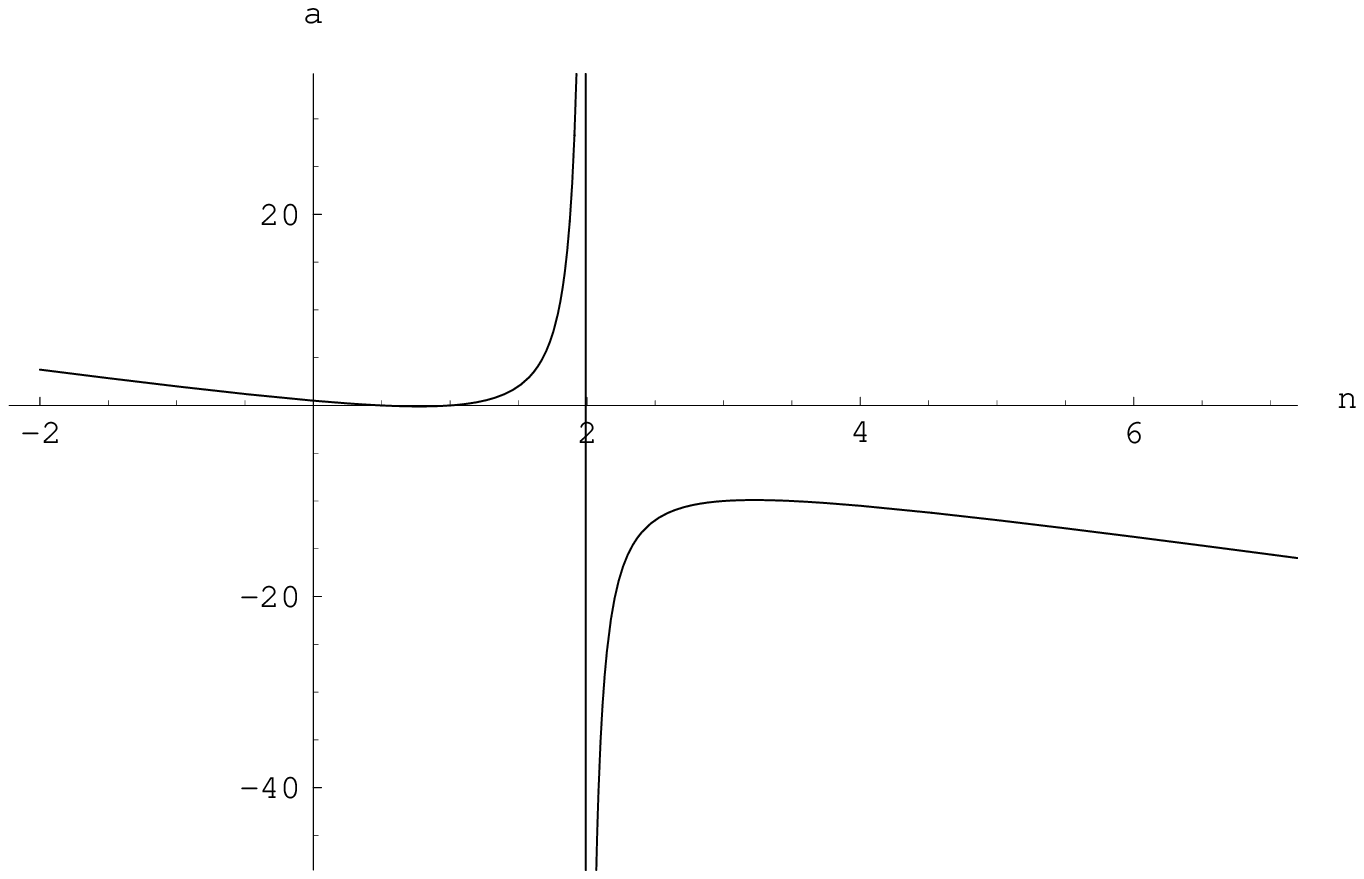}
  \includegraphics[width=5cm]{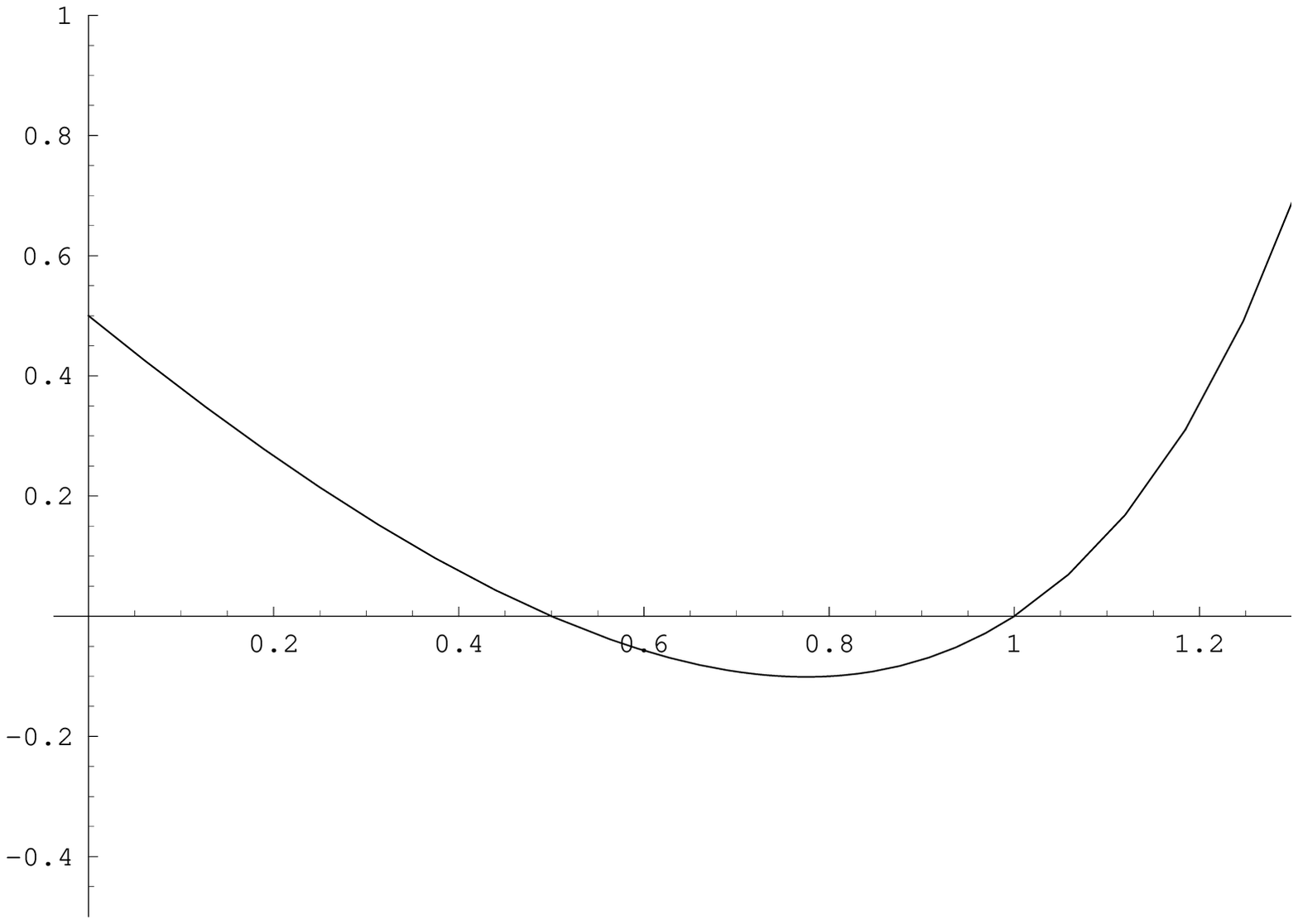}
  \\
 \caption{\small The behaviour
of $\alpha$ in term of $n$. It is evident a region in which the
power of the scale factor is more than one and a region in which
it is always negative. The plot on the right represent the values
of $n$ between 0 and 1.5.}\label{fig1}
\end{center}
\end{figure*}

Using Eqs.(\ref{18}) and (\ref{19}) we can also deduce
 the state equation (the barotropic index)  for the family
 of solution $\alpha=\displaystyle\frac{2n^2-3n+1}{2-n}$. We have
\vspace{7mm}
\begin{equation}\label{29}
w_{(curv)}=-\left(\frac{6n^2-7n-1}{6n^2-9n+3}\right)\,,
\end{equation}
which clearly is $w_{(curv)}\rightarrow{-1}$ for $n\rightarrow
{\infty}$. This fact shows that the approach is compatible with
the recovering of a cosmological constant.
\\
\begin{figure}
\begin{center}
    \includegraphics[width=12cm]{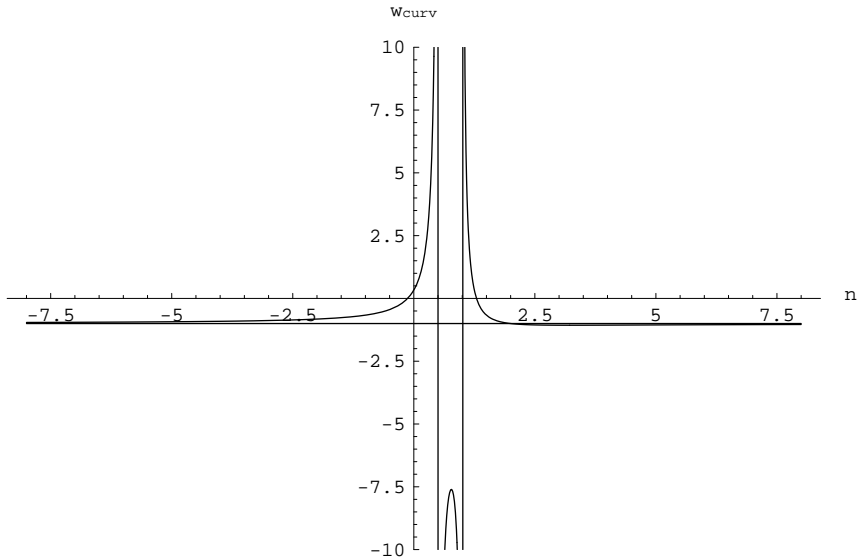}\\
  \caption{\small Behaviour of $w_{curv}$ against $n$.
We have drawn a line for the value $w_{curv}=$$-1$ emphasizing the
cosmological constant state equation value.} \label{2}
\end{center}
\end{figure}

\noindent Eq.(\ref{29}) is plotted in Fig.2. From the two plots,
we can observe that the accelerated behaviour is allowed only for
$w_{(curv)}< 0$ as
requested for a cosmological fluid with negative pressure. \\
The whole approach seems intriguing. In fact we are able to
describe the accelerated phase of universe
 expansion simply as an effect of higher order curvature terms which provide
 an effective negative pressure contribute. In order to see if
 such behaviour is possible for today epoch we have to match our model with observational data.

\section{Matching with SNIa observations}

To verify if the curvature quintessence approach is an interesting
perspective, we have to match the model with the observational
data. In this way, we can constrain the parameters of the theory
to significant values. First we compare our theoretical setting
with the SNeIa results. As a further analysis, we check also the
capability of our model in the universe
age prediction.\\
We have stressed in Sec.1 that SNeIa observations have represented
a cornerstone in the recent cosmology, pointing out that we live
in an expanding accelerating universe. This result has been
possible in relation to the feature of supernovae to be considered
standard candles via the {\it Phillips amplitude-luminosity
relation}.\\
To test our cosmological model, we consider supernovae
observations reported in \cite{perlmutter} \cite{riess}. We have
compiled a combined sample of 79 SNeIa discarding 6 likely
outliers SNeIa as
discussed in \cite{perlmutter}.\\
Starting from these data, it is possible to perform a comparison
between the theoretical expression of the distance modulus

\begin{equation} \mu(z) = 5 \log{\frac{c}{H_0} d_L(z)} + 25\,,
\label{30}
\end{equation}

\noindent and its experimental value for SNeIa, $z$ is the
redshift ($c$ is the light speed, hereafter we will use standard
units). In general, the luminosity distance $d_{L}$ can be
expressed as:

\begin{equation}\label{31}
d_{z}=\frac{c}{H_{0}}(1+z)\int_{0}^{z}{\frac{1}{E({\zeta})}}d{\zeta}\,,
\end{equation}
\vspace{7mm}

\noindent where $E(\zeta)=\displaystyle\frac{H}{H_{0}}$. From
(\ref{22}), the Hubble parameter is

\begin{equation}\label{32}
H(z)=H_{0}(1+z)^{1/{\alpha}}\, ,
\end{equation}

\vspace{7mm}

\noindent where $\alpha$ depends on $n$ as in Eq.(\ref{28}). A fit
by SNeIa data provides significant values of the parameter $n$
which assigns the curvature function.
\\ The best fit is performed as in \cite{wang} minimizing the
$\chi^2$ calculated between the theoretical and the observational
value of distance modulus:

\begin{equation}
\chi^2(H_0, n) = \sum_{i}{\frac{[\mu_i^{theor}(z_i | H_0, n) -
\mu_i^{obs}]^2} {\sigma_ {\mu_{0},i}^{2} + \sigma_{mz,i}^{2}}}
\label{33}
\end{equation}

\vspace{7mm}

\noindent for the above considered data \cite{perlmutter,riess}.
\\
Now, the luminosity distance becomes:

\begin{equation}\label{34}
d_{L}(z,H_{0},n)=\frac{c}{H_{0}}(1+z)\int_{0}^{z}(1+\zeta)^{-1/\alpha}d{\zeta}
\end{equation}
\\
which, after integration, gives

\begin{equation}\label{35}
d_{L}(z,H_{0},n)=\frac{c}{H_{0}}\left(\frac{\alpha}{\alpha-1}\right)(1+z)\left[(1+z)^{\frac{\alpha}{\alpha-1}}-1\right].
\end{equation}
\vspace{7mm}

\noindent This expression is not defined for $\alpha=0,1$ which
physically correspond to static universes and Milne ones; such
models are not interesting for our purposes. The range of $n$ can
be divided into intervals taking into account the existence of
singularities in (\ref{35}). Thus, the fit is performed in five
intervals of $n$, which are:
$n<\displaystyle\frac{1}{2}(1-\sqrt{3})\,,\,
\frac{1}{2}(1-\sqrt{3})<n<\frac{1}{2}\,$,\\$\,\displaystyle
\frac{1}{2}<n<1$ \,,
$\displaystyle1<n<\frac{1}{2}(1+\sqrt{3})\,$\,,
$\,\displaystyle n>\frac{1}{2}(1+\sqrt{3})$.\\
In order to define a limit for $H_{0}$, we have to note that the
Hubble parameter, as a function of $n$, has the same trend of
$\alpha$ showed in Fig.1. We find that for $n$ lower than --100,
the trend is strictly increasing while for $n$ positive, greater
than 100, it is strictly decreasing. In relation to this feature,
we have tested the values of $n$ ranging in these limits because,
as we shall see below, out of this range, the value of the age of
universe becomes manifestly non-physically significant. The
results of the fit are showed in Table 1. In Fig.3, we present the
different contour plots for each evaluated range.\\

\vspace{1cm}
\begin{table}
\begin{center}
 \begin{tabular}{|c|c|c|c|}
  \hline
    Range & $H_{0}^{best}$($km\,s^{-1}Mpc^{-1}$)& $n^{best}$& $\chi^{2}$ \\
  \hline
  $-100<n<1/2(1-\sqrt{3})$ & $65$ & $-0.73$ & $1.003$  \\ \hline
  $1/2(1-\sqrt{3})<n<1/2$ & $63$ & $-0.36$ & $1.160$ \\ \hline
  $1/2<n<1$ & $100$ & $0.78$ & $348.97$ \\ \hline
  $1<n<1/2(1+\sqrt{3})$ & $62$ & $1.36$ & $1.182$ \\ \hline
  $1/2(1+\sqrt{3})<n<3$ & $65$ & $1.45$ & $1.003$ \\ \hline
  $3<n<100$ & $70$ & $100$ & $1.418$ \\ \hline
\end{tabular}
\end{center}
\caption{\small Results obtained by fitting the curvature
quintessence models against SNeIa data. First column indicates the
range of $n$, column two gives the relative best fit value of
$H_{0}$, column three $n^{best}$, column four the $\chi^{2}$
index.}
\end{table}

\begin{figure*}
\begin{center}
\includegraphics[width=6cm]{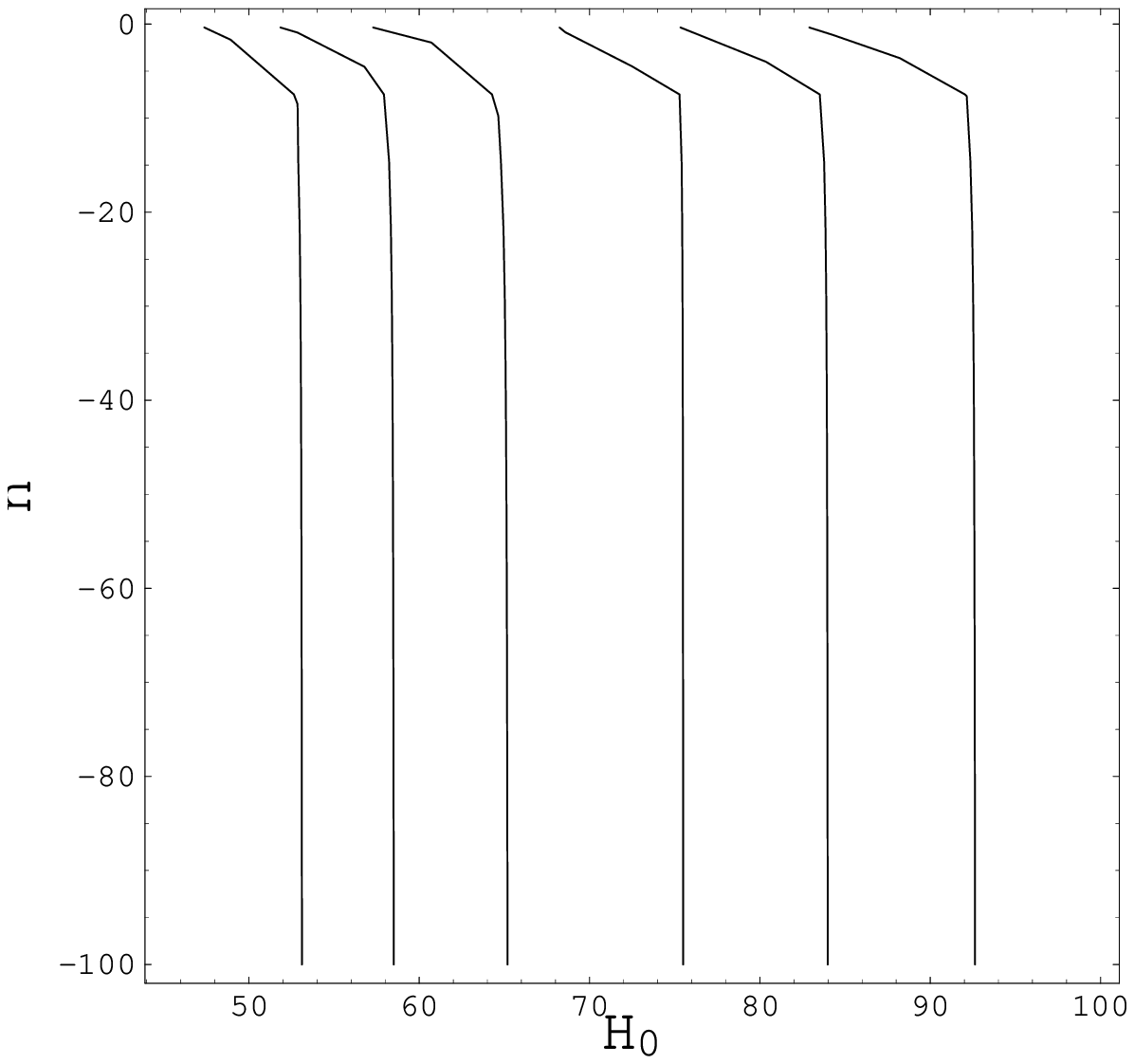}
\includegraphics[width=6cm]{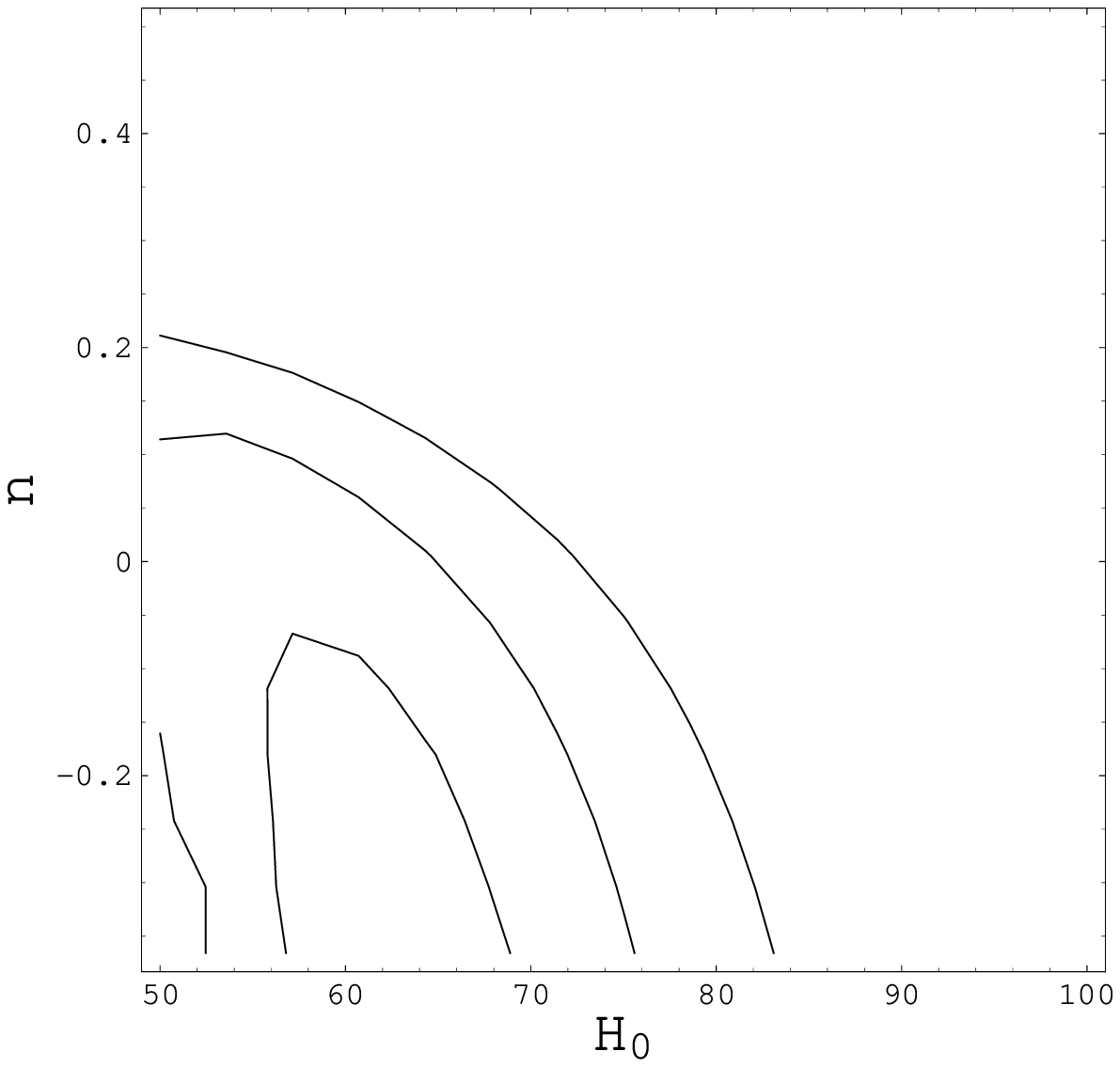}
\includegraphics[width=6cm]{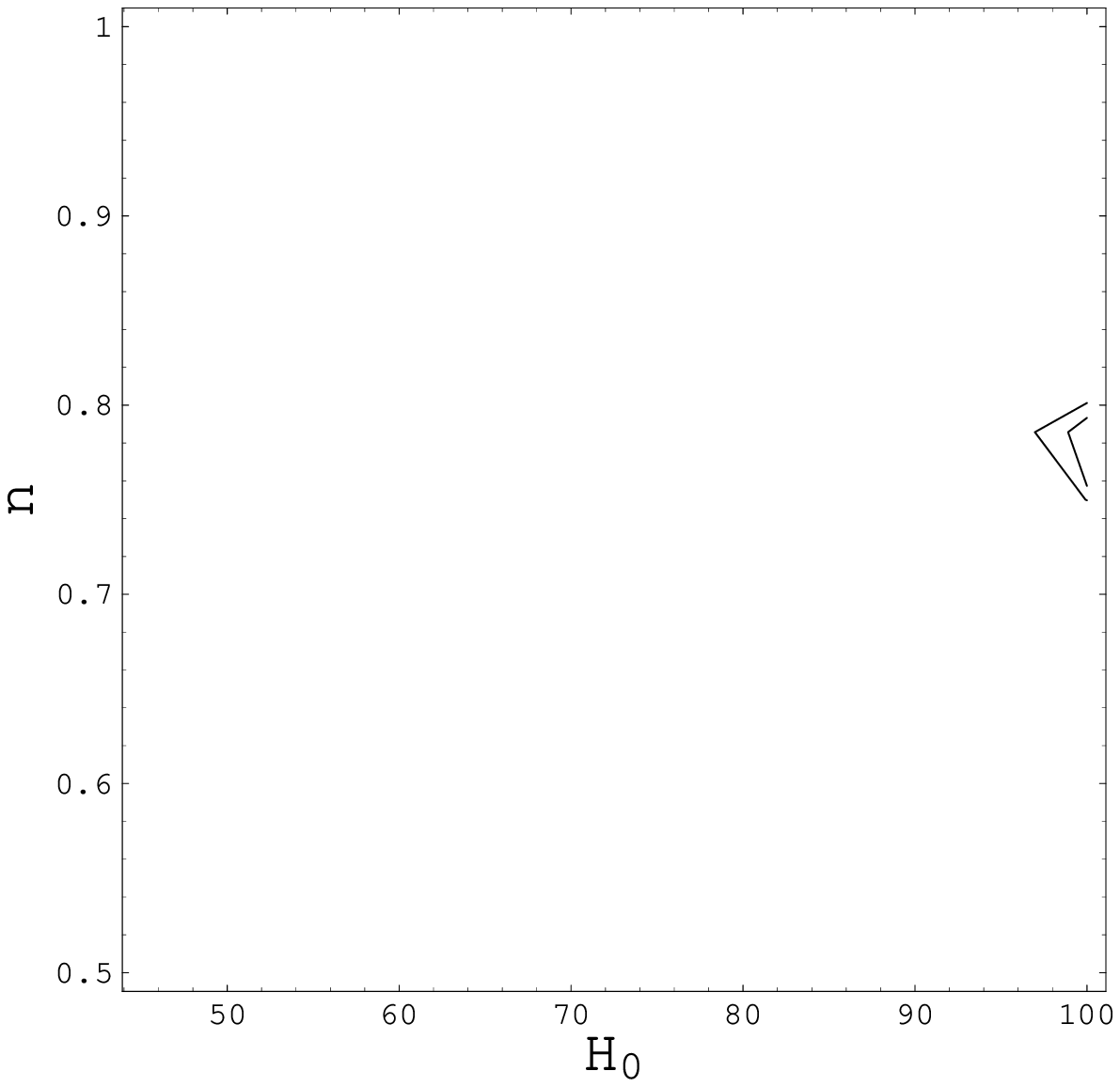}
\includegraphics[width=6cm]{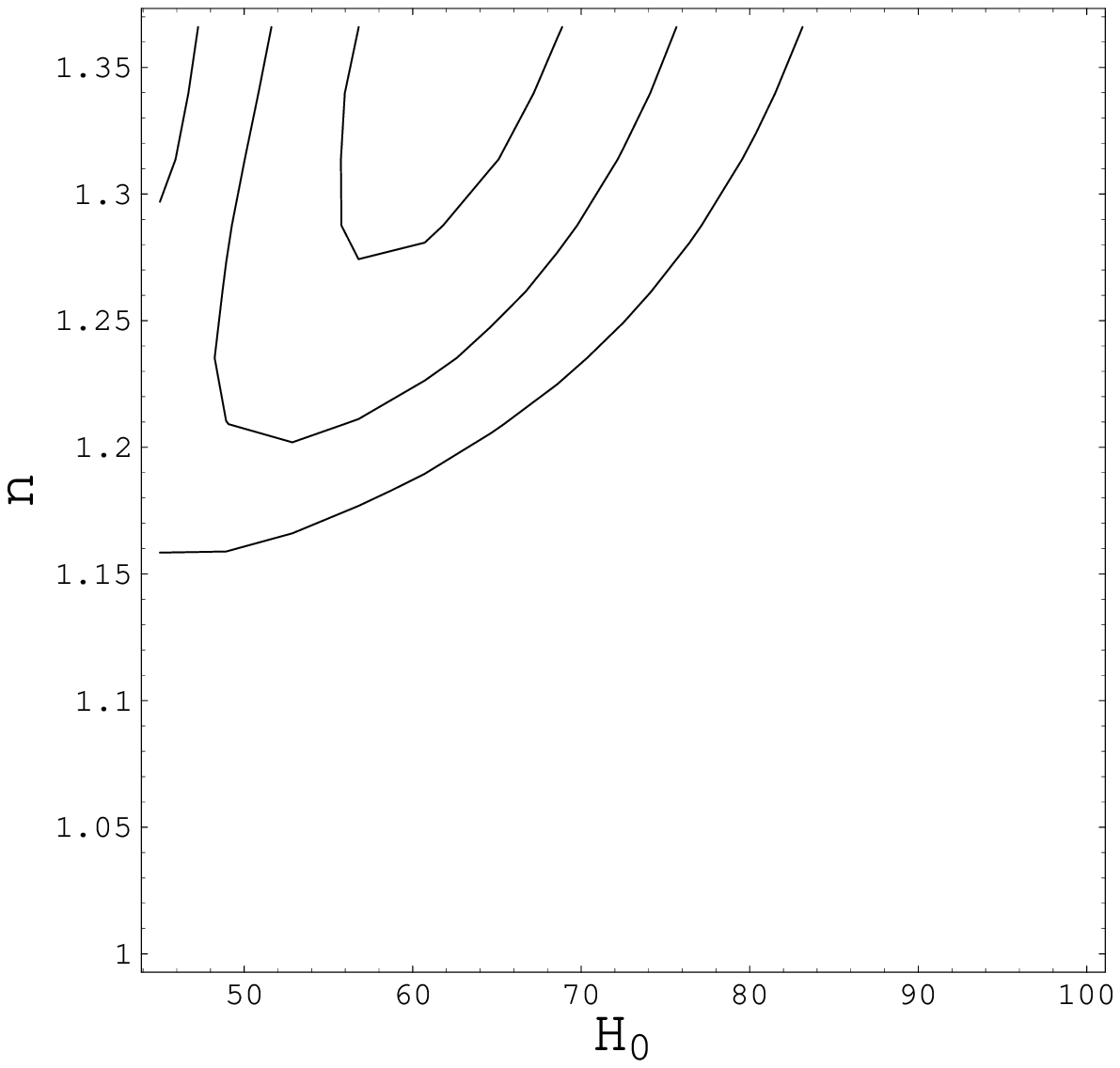}
\includegraphics[width=6cm]{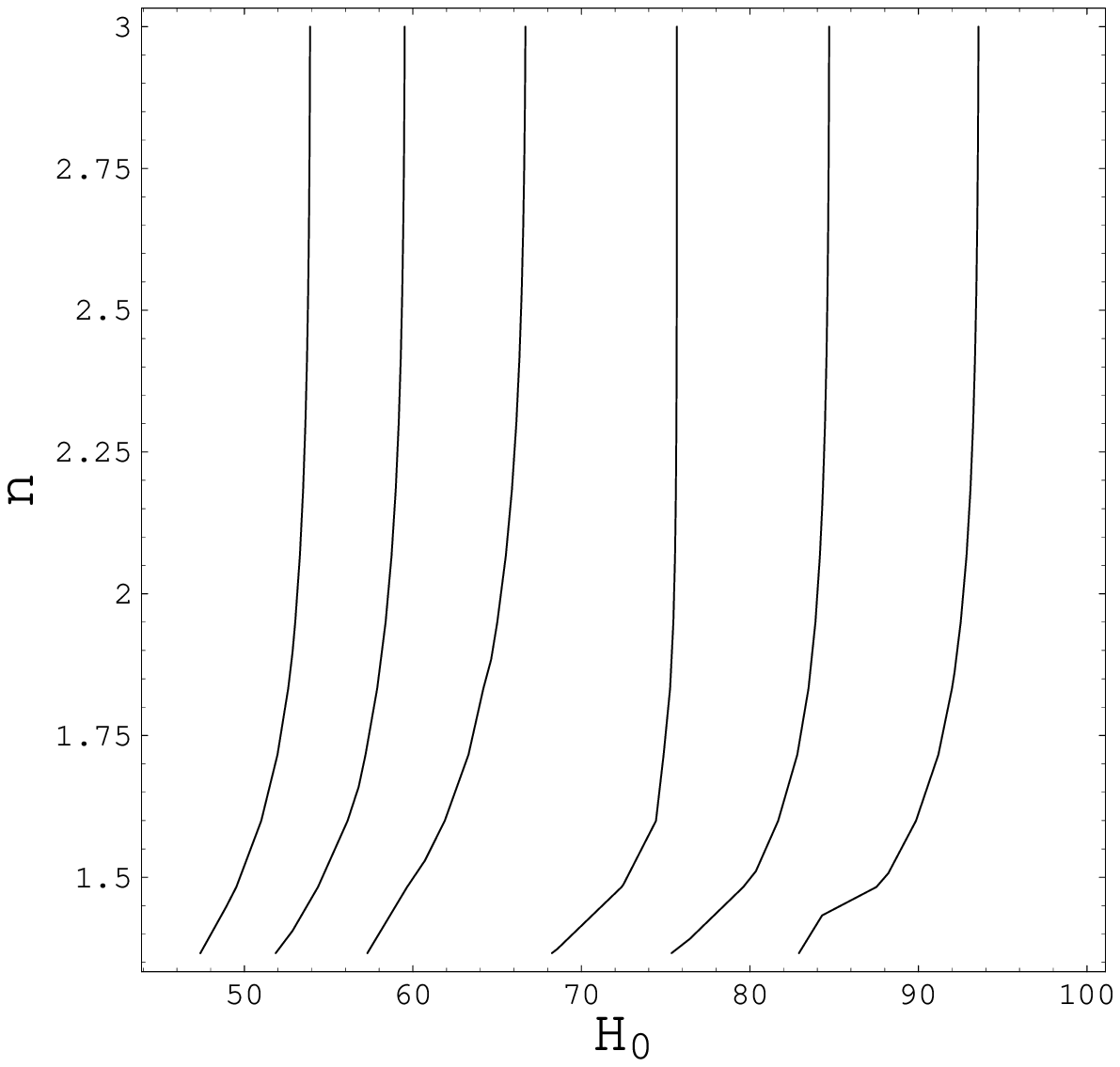}
\includegraphics[width=6cm]{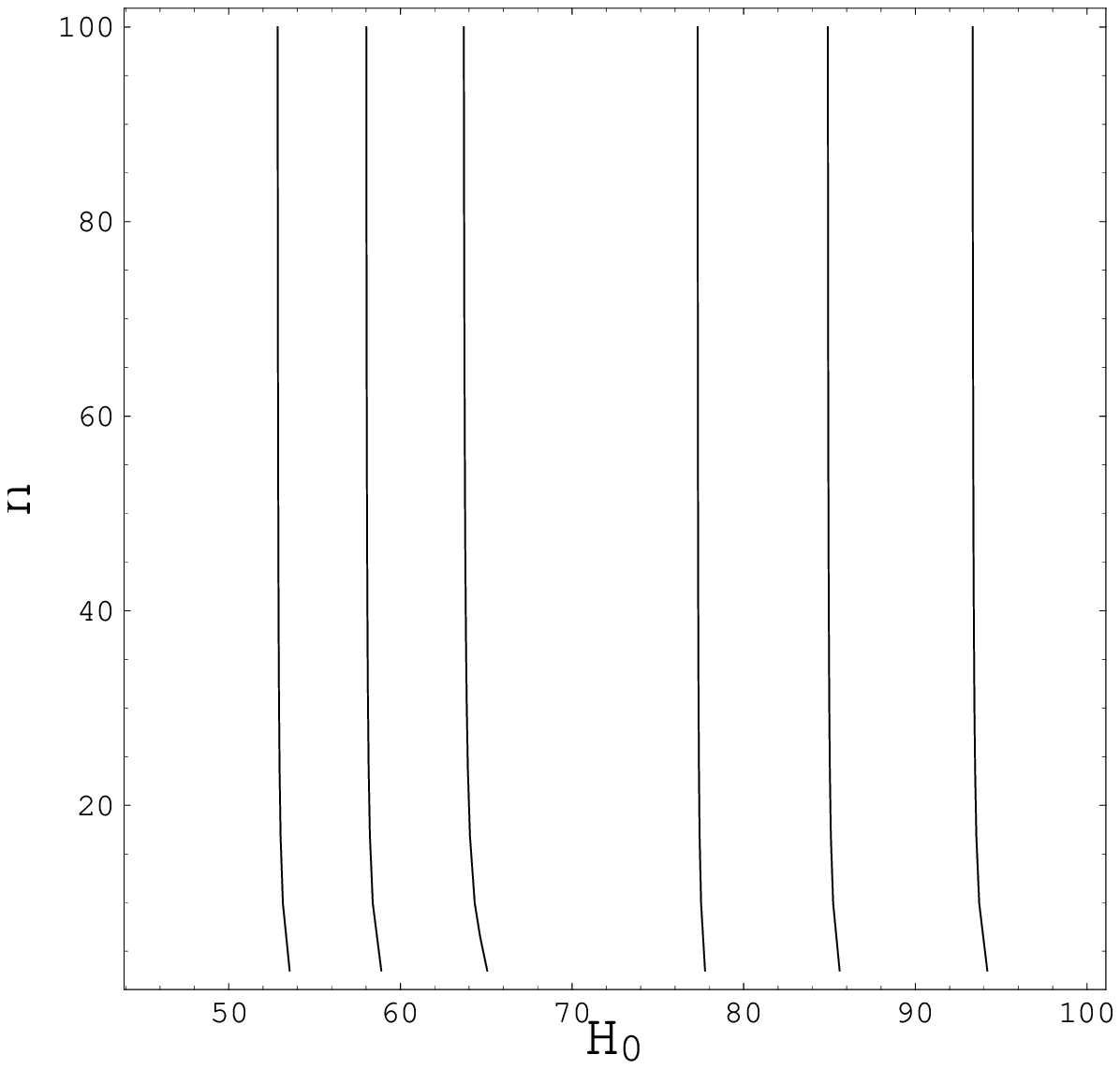}
\caption{\small Contour plots for the considered $n$-ranges using
the SNIa data. It is evident that in the interval $1/2<n<1$ the
models are non physical. The ranges $1/2(1-\sqrt{3})<n<1/2$ and
$1/2<n<1$ give some indications on the best fit value of $n$, in
the other cases the matching with SNIa is completely degenerate.}
\end{center}
\end{figure*}

\noindent From Tab.1 and Fig.3, it is evident that only in the
ranges $1/2(1-\sqrt{3})<n<1/2$ and $1/2<n<1$ we can achieve a
constraint on the values of $n$. Besides, we can exclude the range
$1/2<n<1$ as physically not interesting, in relation to the best
fit value of $H_{0}$ and the $\chi^{2}$ result. In the other
cases, the test gives significant best-fit values both for Hubble
parameter and $\chi^{2}$. But the contour plots are completely
degenerate with respect to the $n$ parameter. This occurrence
suggests that $\chi^{2}$ varies slightly with respect to $n$ in
this cases, hindering the possibility to constrain $n$.\\
In order to discriminate the significant value of $n$, we have
also performed a test of the model in relation to the capability
of estimating the age of the universe.\\ \\
The age of the universe can be simply obtained, from a theoretical
point of view, if one knows the value of the Hubble parameter. In
our case, from the definition of $H$, and using the relations
(\ref{22}), we have:

\begin{equation}
\label{36} t=\alpha H^{-1}
\end{equation}

\vspace{7mm}

\noindent which is

\begin{equation}\label{37}
t=\left(\frac{2n^2-3n+1}{2-n}\right)H^{-1} .
\end{equation}
\vspace{7mm}

\noindent The age of universe can be obtained substituting
$H=H_{0}$ in (\ref{37}).

\noindent We evaluate the age taking into account the intervals of
$n$ and the $3\sigma$-range of variability of the Hubble parameter
deduced by the Supernovae fit. We have considered, as good
predictions, age estimates
included between 10$Gyr$ and 18$Gyr$. \\ \\

\noindent By this test, we are able of refine the allowed values
of $n$. The results are shown in Table 2. First of all, we discard
the intervals of $n$ which give negative values of $t$. Eq.
(\ref{36}) shows that negative values of the age of the universe
are obtained for negative values of $\alpha$, so we have to
exclude the range $1/2<n<1$ and $n>2$ (Fig.1). Conversely, the
other ranges, tested by SNIa fit (Tab.1), become narrower,
strongly constraining $n$.

\begin{table}
\begin{center}
 \begin{tabular}{|c|c|c|c|}
  \hline
  $Range$ & $\Delta H(km\,s^{-1}Mpc^{-1})$& $\Delta{n}$ & $t(n^{best})(Gyr)$ \\
  \hline
  $-100<n<1/2(1-\sqrt{3})$& $50-80$ & $-0.67\leq n<-0.37$ & $23.4$ \\ \hline
  $1/2(1-\sqrt{3})<n<1/2$ & $57-69$ & $-0.37<n \leq -0.07$ & $15.6$ \\ \hline
  $1<n<1/2(1+\sqrt{3})$ & $56-70$ & $1.28\leq n<1.36$ & $15.3$ \\ \hline
  $1/2(1+\sqrt{3})<n<2$ & $54-78$ & $1.37<n\leq 1.43$ & $24.6$ \\ \hline
\end{tabular}
\end{center}
\caption{\small The results of the age test. In the first column
is presented the tested range. Second column shows the
$3\sigma$-range for $H_{0}$ obtained by supernovae test, while in
the third we give the $n$ intervals, i.e. the values of $n$ which
allow to obtain ages of the universe ranging between 10$Gyr$ and
18 $Gyr$. In the last column, the best fit age values of each
interval are reported.}
\end{table}
\vspace{1cm} \noindent A further check for the allowed values of
$n$ is to verify if the interesting ranges of $n$ provide also
accelerated expansion rates. This test can be easily performed
considering the definition of the deceleration parameter
$\displaystyle
q_{0}=-\left.\left(\frac{{\ddot{a}}a}{{\dot{a}^2}}\right)\right|_{\,0}$,
using the relation (\ref{22}) and the definition of $\alpha$ in
term of $n$. To obtain an accelerated expanding behaviour, the
scale factor $a(t)=a_{0}t^{\alpha}$ has to get negative or
positive values of $\alpha$ greater than one. We obtain is that
only the intervals $-0.67\leq n \leq 0.37$ and $1.37\leq n \leq
1.43$ provide a negative deceleration parameter with $\alpha>1$.
Conversely the other two intervals of Tab.2 do not give
interesting cosmological
dynamics, being $q_{0}>0$ and $0<\alpha<1$.\\

\section{WMAP Age test}

A further test of the model can be performed by the age estimate
obtained by the WMAP campaign. The WMAP ({\it Wilkinson Microwave
Anisotropy Probe}) mission aims to determine the geometry, the
content and the evolution of the universe through a full sky map
of the temperature anisotropies of cosmic microwave background
radiation \cite{wmap}. The first year observational results has
been published since few months. These observations indicate as
the best fit model a $\Lambda$-term cosmological model with about
$70\%$-content of cosmological origin, Hubble constant value
$71^{+0.04}_{-0.03}\, km \, s^{-1}Mpc^{-1}$ and an age estimate of $13.7^{+0.2}_{-0.2}Gyr$.\\
Using this last result, we can improve the constraints on $n$ in
relation to the very low error ($1\%$) of WMAP age
estimator.\\
We use the same approach of the previous section, the only
difference is to consider as physically interesting only the age
prediction ranging between $13.5Gyr$ and
$13.9Gyr$. The results are shown in Tab.3. \\
It is evident that this test narrows the range of physical
interest. If we take into account also the capability of providing
acceleration, we obtain that the interesting
values of $n$ are $-0.450\leq n<-0.370$ and $1.366<n\leq 1.376$.\\
These results could represent a selection for the allowed form of
fourth-order gravity action as a power law of curvature Ricci
scalar.

\begin{table}[h!]
\begin{center}
 \begin{tabular}{|c|c|c|c|}
  \hline
   $Range$ & $\Delta H(km\,s^{-1}Mpc^{-1})$& $\Delta{n}$ & $q_{0}$ \\
  \hline
  $-100<n<1/2(1-\sqrt{3})$& $50-80$ & $-0.450\leq n<-0.370$ & $<0$ \\ \hline
  $1/2(1-\sqrt{3})<n<1/2$ & $57-69$ & $-0.345<n \leq -0.225$ & $>0$ \\ \hline
  $1<n<1/2(1+\sqrt{3})$ & $56-70$ & $1.330\leq n<1.360$ & $>0$ \\ \hline
  $1/2(1+\sqrt{3})<n<2$ & $54-78$ & $1.366<n\leq 1.376$ & $<0$ \\ \hline
\end{tabular}
\end{center}
\caption{\small The results of the age test for the curvature
quintessence model based on the prediction of WMAP observations.
In the last column we show the sign of the deceleration parameter
as a testify of the potential accelerating rate of the model.}
\end{table}

\section{Conclusions}

 In this paper we have analyzed a geometrical approach to
 quintessence given by considering a fourth-order theory of gravity and we have
 matched the cosmological models derived with observational data. This scheme,
 proposed as {\it curvature quintessence} \cite{capoz-quint curv}, represents an approach
 to describe an accelerated expanding universe dominated by a
 cosmological component without using scalar fields.\\
 We stress that such an approach has a natural background in
 several attempts of quantize gravity, because higher-order curvature invariants
 are generated to renormalize quantum field theories on curved space
 times \cite{buchbinder}. \\ We have studied curvature quintessence
 neglecting matter as a first significative approximation. This
 approximation is possible if we consider that, at one loop level,
 matter and gravity have the same ``weight" in early universe
 \cite{birrell} and today the amount of cosmological and matter
 components, at cosmological density level, are comparable. In both regimes, the overall
 evolution can be assigned only by the cosmological component which gives a good
 approximation of dynamics. We ask for a power law functions for the action (in terms of Ricci
 curvature scalar) and for the scale factor (in terms of
 cosmological time). With these choices, we obtain a family of exact
 solutions so that the model is characterized by one parameter
 (specifically $n$). To check these solutions, we have fitted the model with
observational data. A straightforward test is a comparison with
SNIa observations referring to the well known data of SCP ({\it
Supernovae Cosmology Project} \cite{perlmutter}) and of HZT
({\it High-Z search Team} \cite{riess}).\\
The model fits these data and provides a constrain on the
parameter $n$. Unfortunately this test is not decisive to feature
curvature quintessence since the likelihood curves are
degenerate with respect to $n$.\\
To improve this result, we have performed a test with the age of
the universe.\\
In a first case, our test has been conceived considering, as good
estimates for age of universe, values ranging between 10$Gyr$ and
18 $Gyr$. Considering that we take into account only accelerated
dynamics, we deduced that for $-0.67\leq n \leq 0.37$ and
$1.37\leq n \leq 1.43$ curvature quintessence is a model capable
of mimicking
the actual universe.\\
In order to better refine these ranges, we have then considered a
test based on WMAP age evaluation.\\
In this case, the age ranges between $13.5Gyr$ and $13.9Gyr$. This
fact reduces the allowed intervals of the parameter $n$, which are
$-0.450\le n <-0.370$ and $1.366<n<1.376$. In conclusion, we can
say that a fourth order theory of gravity of the form

\vspace{4mm}
\begin{equation}
f(R)=f_{0}R^{1+\varepsilon}
\end{equation}
\vspace{4mm}

\noindent with $\varepsilon \simeq -0.6$ or $\varepsilon \simeq
0.4$ can give rise to reliable cosmological
models which well fit SNeIa and WMAP data.\\
In this sense, we need only ``small" corrections to second order
Einstein gravity action in order to achieve quintessence issues.\\
Indications in this sense can be found also in a detailed analysis
of $f(R)$ cosmological models performed against CMBR constraints,
as shown in \cite{hwang}.


\begin{thebibliography}{99}


\bibitem{perlmutter}  S. Perlmutter et al.  \apj {\bf 483}, 565 (1997).
    \\ S. Perlmutter et al. {\it Nature} {\bf 391}, 51
    (1998).\\ S. Perlmutter et al. \apj {\bf 517}, 565 (1999).
\bibitem{riess} B.P. Schmidt et al. \apj {\bf 507}, 46
    (1998).\\ A.G. Riess et al. \apj {\bf 116}, 1009
    (1998).
\bibitem{boomerang} P. de Bernardis et al. {\it Nature} {\bf 404}, 955 (2000).
\bibitem{maxima}  A. Balbi et al.\apj {\bf 558}, L145-L146 (2001);
R. Stompor et al. \apj {\bf 561},  L7-L10 (2001).
\bibitem{cobe} G.F. Smoot, SLAC Beam Line {\bf 23N3} 2, (1993);
 C. L. Bennet et al. \apj {\bf 464}, L1 (1996);
A. H. Jaffe et al \prl {\bf 86}, 3475 (2000).
\bibitem{sunyaev} Y. Repaheli, astro-ph/0211422.
\bibitem{wmap} C.L. Bennet et al. (WMAP collaboration) astro-ph/0302207 (2003).\\ D.N. Spergel et
al.(WMAP collaboration) astro-ph/0302209 (2003).
\bibitem{carroll} S. M. Carroll, Living Rev.Rel. {\bf 4}, 1
(2001).
\bibitem{starobinsky} A. A. Starobinsky, V. Sahni, \ijmp {\bf D9},
373-444(2000).
\bibitem{straumann} N. Straumann, astro-ph/0203330.
\bibitem{steinhardt} R.R. Caldwell,  R. Dave, P.J. Steinhardt,
    \prl {\bf 80}, 1582 (1998).
\bibitem{rubano} R. de Ritis et al., \pr {\bf D 62} (2000) 043506.\\
                 C. Rubano and J.D. Barrow, \pr {\bf D64} (2001)
                 127301.\\
                 C. Rubano and P. Scudellaro, \grg {\bf 34}, 1931 (2002).
\bibitem{sahni} A. Ujjaini, V. Sahni, astro-ph/0203443; V. Sahni,
Y. Shtanov, astro-ph/0202346.
\bibitem{cardassian} K. Freese, M. Lewis, astro-ph/0201229; Y. Wang,
 K. Freese, P. Gondolo, M. Lewis, astro-ph/0302064.
\bibitem{chaplygin}  N. Bilic, G. B. Tupper, R. D. Viollier, \pl
{\bf B535} 17, (2002); J. C. Fabris, S. V. B. Goncalves, P. E. de
Souza, astro-ph/0207430; A. Dev, J. S. Alcanitz, D. Jain, \pr {\bf
D67}, 023515 (2003).
\bibitem{stornaiolo} S. Capozziello, G. Lambiase and C. Stornaiolo, \aph {\bf 10}, 713
(2001).
\bibitem{capoz-torsione} S. Capozziello, \mpl {\bf A17}, 1621
(2002).
\bibitem{quint-torsione} S. Capozziello, V.F. Cardone,
E. Piedipalumbo, M. Sereno, A. Troisi, \ijmp in printing,
astro-ph/0209610.
\bibitem{capoz-quint curv} S. Capozziello, \ijmp {\bf 11}, 483 (2002).
\bibitem{birrell} N.D. Birrell  and P.C.W. Davies  {\it Quantum Fields in Curved
        Space} (1982) Cambridge Univ. Press (Cambridge).
\bibitem{buchbinder} I.L. Buchbinder, S.D. Odintsov, I.L. Shapiro,
 {\it Effective Action in Quantum Gravity} IOP Publishing (1992)
Bristol.
\bibitem{starobinsky80} A.A. Starobinsky, \pl {\bf B 91} (1980) 99.
\bibitem{kerner1} R. Kerner, \grg {\bf 14}, 453, (1982).
\bibitem{kerner2} J. P. Duruisseau, R. Kerner, \cqg {\bf 3}, 817,
(1986).
\bibitem{pavon} A. B. Balakin, D. Pavon, D. J. Schwartz, W. Zimdahl, astro-ph/0302150.
\bibitem{lambda} S. Capozziello, R. de Ritis, and A.A. Marino  {\it Gen. Rel. Grav.}
                 {\bf 30} (1998) 1247.
\bibitem{capoz-lambiase} S. Capozziello, G. Lambiase, \grg {\bf 32}, 295
(2000).
\bibitem{quartic} S. Capozziello, R. de Ritis, C. Rubano, P. Scudellaro, \ijmp
        {\bf D 4} (1995) 767.
\bibitem{wang}
Y. Wang, \apj {\bf 536}, 531 (2000).
\bibitem{hwang} J. Hwang, H. Noh, \pl {\bf 506 B}  (2001) 13.

\end{thebibliography}
\end{document}